\documentclass[prl, superscriptaddress, showpacs, floatfix,
twocolumn]{revtex4}

\usepackage{oldgerm}
\usepackage{latexsym}
\usepackage{amsmath}
\usepackage{amsgen}
\usepackage{amscd} 
\usepackage{color}
\usepackage{epsfig}
\usepackage{amsfonts}

\input{diagrams}
\diagramstyle[height=8mm,width=1.4cm]
\newarrow{Bond}	----{>}

\newtheorem{thm}{Theorem}
\newtheorem{lem}{Lemma}
\newtheorem{prop}{Proposition}

\newtheorem{dfn}{Definition}
\newcommand{\opr}[1]{\operatorname{#1}}
\newenvironment{proof}{\noindent{\it Proof.}} {\hfill $\Box$

}

\begin{document}

\title{Quantum Computer Condition: Stability, Classical Computation and Norms} \author{Gerald Gilbert, Michael Hamrick, F. Javier Thayer and Yaakov S. Weinstein\\
\small \it Quantum Information Science Group\\
{\sc  Mitre} \\
\small \it 260 Industrial Way West, Eatontown, NJ 07724 USA \\
\tt{E-mail: \{ggilbert, mhamrick, jt, weinstein\}@mitre.org}}

\begin{abstract}
The Quantum Computer Condition (QCC) provides a rigorous and completely general framework for carrying out analyses of questions pertaining to fault-tolerance in quantum computers.
In this paper we apply the QCC to the problem of fluctuations and systematic errors in the values of characteristic parameters in realistic systems. We show that fault-tolerant quantum computation is possible despite variations in these parameters. We also use the QCC to explicitly show that reliable classical computation can be carried out using as input the results of fault-tolerant, but imperfect, quantum computation. Finally, we consider the advantages and disadvantages of the superoperator and diamond norms in connection with application of the QCC to various quantum information-theoretic problems. 
\end{abstract}


\maketitle

\section{Introduction}

The Quantum Computer Condition (QCC)~\cite{GHT2005} is a rigorous mathematical statement that connects the irreversible dynamics of the quantum computing machine, with the reversible operations that comprise the quantum computation intended to be carried out by the quantum
computing machine. A discussion of several physical consequences of the QCC is found in~\cite{GHT2005}, including the Quantum Computing No-Go Theorem, which establishes a bound for decoherence and dissipation beyond which quantum computation is not possible.

The quantum computer condition is denoted by the symbol
$\mathbf{QCC}(P,U,\mathcal{M}_{\{\mathrm{l}\rightarrow\mathrm{c}\}} 
\mathcal{M}_{\{\mathrm{c}\rightarrow\mathrm{l}\}},\alpha)$. This holds if and only if, 
for all density matrices $\rho \in \mathbf{T}(H_\mathrm{logical})$, we
have
\begin{equation}\label{encoded_QCC}
\|\mathcal{M}_{\{\mathrm{c}\rightarrow\mathrm{l}\}}(P\cdot(\mathcal{M}_{\{\mathrm{l}\rightarrow\mathrm{c}\}}(\rho))) -
U\rho U^\dag\|_1 \leq \alpha ~.
\end{equation}
The above expression relates a unitary operator $U$ on a
Hilbert space $H_\mathrm{logical}$ of logical qubits with a completely
positive map $P$ on a computational Hilbert space $H_\mathrm{comp}$
via the pair of superoperators
$\mathcal{M}_{\{\mathrm{l}\rightarrow\mathrm{c}\}}:\mathbf{T}(H_\mathrm{logical}) \rightarrow
\mathbf{T}(H_\mathrm{comp})$,
$\mathcal{M}_{\{\mathrm{c}\rightarrow\mathrm{l}\}}:\mathbf{T}(H_\mathrm{comp}) \rightarrow
\mathbf{T}(H_\mathrm{logical})$ which are completely positive,
trace-preserving linking maps~\cite{f10}. Here, $\mathbf{T}(H)$ is the Banach space of trace class operators on a Hilbert space $H$ with the Schatten $1$
norm $\| \cdot\|_1$. The relationship between $P$ and $U$ is crucially expressed by the
parameter $\alpha$ which quantifies how well $P$ approximates $U$. The parameter $\alpha$ will be nonzero for any realistic implementation of a quantum computer due to the inevitable presence of noise and the infeasibility of correcting all possible errors~\cite{Preskill,Lidar}.

The QCC addresses a much broader set of problems than error correction alone, by allowing for error processes to act on the state of the system during encoding, recovery and decoding operations as well 
as during the computation itself. Error correction theory alone does not allow for this. Importantly, the quantity $\alpha$, as such, is not even defined in the theory of error correction. The presence of the crucial parameter $\alpha$ highlights the 
difference between the limited scope of error correction and the more general notion of fault tolerant quantum computing: it reflects the inevitable survival of residual errors in {\em any} realistic implementation of a quantum computer. It is fault-tolerance, and not merely error correction, that is described by the QCC~\cite{CMT}. The advantage of the QCC is that it comprises a rigorously systematic formulation that provides a {\em completely general framework} for carrying out analyses of questions pertaining to fault-tolerance in quantum computers, that broadens the scope of previous approaches to the subject~\cite{f11}.

We will find it convenient in this paper to make use of the superoperator (SO) norm defined for arbitrary linear superoperators
$Q:\mathbf{T}(H) \rightarrow \mathbf{T}(H)$ as
\begin{equation}
\|Q\|_\mathrm{SO}^\mathrm{sa} \equiv \sup\{\|Q(\rho)\|_1: \|\rho\|_1 \leq 1
\mbox{ and } \rho = \rho^\dagger\}~,
\end{equation}
where the superscript ``sa" indicates the restriction of the domain of the 
superoperator $Q$ to self-adjoint operators $\rho$. We may re-express the QCC given
in eq.\eqref{encoded_QCC} in terms of the SO norm with the equivalent inequality 
\begin{equation}\label{QCC-in-terms-of-SO}
\|  P^\mathcal{M} - G\|_\mathrm{SO}^\mathrm{sa} \leq \alpha,
\end{equation}
where $P^\mathcal{M} \equiv \mathcal{M}_{\{\mathrm{c}\rightarrow\mathrm{l}\}} P \mathcal{M}_{\{\mathrm{l}\rightarrow\mathrm{c}\}}$ and
$G$ is the map $G(\rho) \equiv  U\rho U^\dagger$. This is the form of the QCC that we will use in the rest of this paper.  

The paper is organized as follows. We consider three applications of the QCC to problems 
arising in quantum computing: (1) we study the stability of quantum computation under variation of characteristic parameters, (2) we analyze the use of realistic, fault-tolerant quantum computation in enabling subsequent classical computation, and (3) we analyze the advantages and disadvantages of the SO and diamond norms as measures of the difference between a desired operation and its physical implementation.

\section{Stability Under Variation of Characteristic Parameters}

Realistic implementations of quantum computers will exhibit fluctuations and systematic errors in the values of characteristic parameters. In this Section we consider the stability of the QCC under small such variations of parameters.  Although the general form of the QCC is valid for either Markovian or non-Markovian underlying dynamics, in this Section of this paper only, we restrict consideration to the case of Markovian dynamics. The system state, $\rho(t)\in \mathbf{T}(H)$, is then governed by an equation of the form 
\begin{equation}
\label{equation-of-motion}
\frac{d}{dt} \rho(t) = A \rho(t),
\end{equation}
where $A$ is a (possibly unbounded) operator on $\mathbf{T}(H)$. In this analysis, we consider time-independent operators $A$, with $A$ otherwise unrestricted. Then, in the sense of analytic semigroup theory,
\begin{equation}
\rho(t) = \exp t A \rho(0)~.
\end{equation}
The propagator, $P=P(t)\equiv\exp tA$, associated to $A$, is a completely positive, trace-preserving map.

We now consider the stability of the QCC under small, time-independent perturbations of
$A$. As an immediate consequence of \eqref{QCC-in-terms-of-SO}, the QCC has the following property: 
\begin{lem}\label{stability-thm}
If $\mathbf{QCC}(P, U, \mathcal{M}_{\{\mathrm{l}\rightarrow\mathrm{c}\}},
\mathcal{M}_{\{\mathrm{c}\rightarrow\mathrm{l}\}}, \alpha)$ holds and $P'$ is a completely
positive trace-preserving map, then $\mathbf{QCC}(P', U,
\mathcal{M}_{\{\mathrm{l}\rightarrow\mathrm{c}\}}, \mathcal{M}_{\{\mathrm{c}\rightarrow\mathrm{l}\}}, \alpha^\prime )$ also holds,
where $\alpha^\prime\equiv\alpha+\|P -P'\|_\mathrm{SO}^\mathrm{sa}$.
\end{lem}
As a consequence of Lemma \ref{stability-thm} we see that replacement of $P$ with $P^\prime$ will result in a quantum computer implementation inaccuracy $\alpha^\prime$ that is close to the original value of $\alpha$ if the difference between $P$ and $P^\prime$ under the (self-adjoint) SO norm is small~\cite{f12}.

Having established Lemma \ref{stability-thm} we may proceed with the analysis of the stability of the QCC under small perturbations of $A$.
Suppose $A$ depends continuously on a parameter $z$, with $z$ taking values
in some topological space. We denote the $z$-dependence of $A$ as $A_z$. The continuity property of such (possibly unbounded) operators, $A_z$,
is defined in terms of a corresponding continuity property of the resolvent
\begin{equation}
\opr{R}(\lambda, A_z) \equiv (\lambda I - A_z)^{-1}.
\end{equation}

We now apply this continuity property of $A_z$ to the stability of the QCC under 
variation of parameters.  From Lemma \ref{stability-thm}, and the observation that convergence of the resolvent
$R(\lambda,A_z)\rightarrow R(\lambda,A_{z^\prime})~\forall\lambda > 0$ implies 
$\exp tA_z\rightarrow\exp tA_{z^\prime}~\forall t>0$, we obtain the following result:

\begin{prop}\label{stability}
Suppose that $\opr{R}(\lambda, A_z)$ is norm operator continuous in $z$ for all $\lambda > 0$.
Then, if for some $t$, $\mathbf{QCC}(\exp t A_z, U, \mathcal{M}_{\{\mathrm{l}\rightarrow\mathrm{c}\}},
\mathcal{M}_{\{\mathrm{c}\rightarrow\mathrm{l}\}}, \alpha)$ holds, then for any
$\alpha'>\alpha$,
$\mathbf{QCC}(\exp t A_w, U, \mathcal{M}_{\{\mathrm{l}\rightarrow\mathrm{c}\}},
\mathcal{M}_{\{\mathrm{c}\rightarrow\mathrm{l}\}}, \alpha')$ holds for $w$ sufficiently near $z$.
\end{prop}

The guarantee of stability under variations in the characteristics of
the dynamics is vital when studying actual experimental
implementations. In practice it is effectively impossible to guarantee the perfect stability of an experimental system. However, because of the QCC stability condition
exhibited in Proposition \ref{stability}, we are nevertheless
guaranteed successful quantum computation, albeit with a possibly larger implementation
inaccuracy ({\em {i.e.}}, the replacement $\alpha\rightarrow\alpha^\prime$). An example of instability in an experiment is the drift of the 
magnetic field in nuclear magnetic resonance. This causes inhomogeneity in the 
quantizing field, which in turn affects the decoherence of 
the system. Another example of instability in controlling experimental parameters arises due to the necessity of turning on and off couplings between quantum dots. The coupling
depends exponentially on distance between the electrons in the dots
and is nearly impossible to control precisely or repeat exactly \cite{Friesen}. 
The stability property of Proposition  \ref{stability} is essential to show that quantum computation is still possible in spite of these effects.

\section{The QCC, Quantum Computation and Classical Computation}

The generic problem to which a quantum computer will be applied will not consist solely of quantum computation {\em per se}, but rather will be comprised of an initial quantum computation followed by a classical computation.
Typically the output of the quantum computation will be utilized as input to the classical computation.
Since quantum computation results in a probabilistically distributed set of outcomes, the question arises as to
whether or not {\em reliable} classical computation can follow. An affirmative answer to this foundational question is necessary in order that algorithms such as Shor's algorithm can be applied to practical problems. A preliminary version of this foundational question is addressed in the
quantum computational model formalized by Kitaev~(\cite{kitaev1997}, \S4.1).
In Kitaev's model, the initial, purely quantum mechanical computation that precedes the subsequent classical computation, is assumed to be perfectly executed with no residual errors~\cite{f13}. In this Section of our paper, we extend and complete the analysis of this foundational question by considering the realistic case in which the physical implementation of the quantum computation includes residual errors. Residual errors are always present, irrespective of the existence of decoherence-free subspaces or noiseless sub-systems, since a non-vanishing residual error probability persists even after error correction is applied~\cite{Lidar,Preskill}. Our extension of Kitaev's model demonstrates that reliable classical computation can follow from fault-tolerant quantum computation.

We begin by reviewing the Kitaev model~\cite{kitaev1997}. This model relates: (1) each instance of a probabilistic {\em classical} computational problem with (2) a {\em quantum} computational circuit which is polynomial time computable as a function of the instance size. This relationship can be represented by a diagram:
\begin{equation} \label{commutative-diagram}
\begin{diagram}
H_\mathrm{logical}  & \rBond^{U} & H_\mathrm{logical} \\
\uBond<{\opr{I_{p}}} & \boxed{p}  &  \dBond>{\opr{O_{p}}} \\
X & \rBond_{F} & Y \end{diagram}~.
\end{equation}
This diagram expresses the fact that the output of the quantum computation 
$U$ is intended to be used in computing the classical function $F$,  where $F:X \rightarrow Y$ is an instance of the classical computational problem.  
Here ${\opr{I_\mathrm{p}}}$ is an initialization map, which maps the classical input
space $X$ into pure states, and $\opr{O_\mathrm{p}}$ is the corresponding readout map, which maps the output of the operation $U$ onto the classical output space $Y$. In general $\opr{O_\mathrm{p}}$ is  a quantum measurement given by a projection-valued measure (or more generally a POVM) $\{\opr{E}_y\}_{y \in
Y}$. The symbol in the center of the diagram refers to the probabilistic inaccuracy, $p$, associated to the output of the classical computation. The quantity $1-p$ is a measure of the probability of success of the final classical computation, for which the quantum computation provides input data~\cite{f14}. The initialization map and readout maps have to be sufficiently simple (polynomial time in the input size) so that they do not implicitly hide complexity. Kitaev's formulation \cite{kitaev1997} requires that the diagram~\eqref{commutative-diagram} be {\em nearly commutative} in a probabilistic sense which we now make precise.

To this end, first replace the diagram in eq.\eqref{commutative-diagram} (which directly makes use of pure states) with the following diagram that instead makes use of density matrices to represent the pure states:
\begin{equation} \label{commutative-diagram_mixed}
\begin{diagram}
{\mathbf{T}}(H_\mathrm{logical})  & \rBond^{G} & {\mathbf{T}}(H_\mathrm{logical}) \\
\uBond<{\opr{I_{\mathrm{m}}}} & \boxed{p}  &  \dBond>{\opr{O_\mathrm{m}}} \\
X & \rBond_{F} & Y \end{diagram}~,
\end{equation}
where the action of the map $G$ is given by $G: \rho \mapsto U \rho U^\dagger$,
the action of the map $\opr{I_\mathrm{m}}$ is given by
$\opr{I_\mathrm{m}}: x \mapsto | \opr{I}_{\mathrm{p}}(x) \rangle \langle \opr{I}_{\mathrm{p}}(x) |$, and $\opr{O}_{\mathrm{m}}$ is the quantum measurement corresponding to $\opr{O}_{\mathrm{p}}$, except that $\opr{O}_{\mathrm{m}}$ acts on density matrices, whereas $\opr{O}_{\mathrm{p}}$ acts directly on vectors contained in $H_\mathrm{logical}$. Given an input $x \in X$, the output of the quantum mechanical computation is distributed according the probability
measure on $Y$ as follows: the probability $\opr{Pr}_x(y)$ of a
singleton $y \in Y$ is $\opr{tr}(\sqrt{\opr{E}_y} U \opr{I}_\mathbf{m}(x) U^\dagger
\sqrt{\opr{E}_y})$. Then, the {\em near commutativity} of the diagram in \eqref{commutative-diagram}
(as well as of the diagram in \eqref{commutative-diagram_mixed}) means that for each $x \in X$, the probability
measure $\opr{Pr}_x(y)$ is sufficiently concentrated at $F(x)$ so that
a majority vote algorithm determines the correct value $F(x)$.  In the
case the output space $Y$ is binary, it suffices there is a $p < 1/2$,
such that
\begin{equation}\label{approximation-cond}
\opr{tr}\left(\sqrt{\opr{E}_{F(x)}} U \opr{I}_\mathbf{m}(x) U^\dagger \sqrt{\opr{E}_{F(x)}}~\right) > 1 - p
\end{equation}
for all $x \in X$. Majority voting will yield a correct result for the classical computation provided that $p$ is sufficiently small ({\em e.g.}, $p < 1/2$ in the case of $Y$ binary). This concludes our review of the Kitaev model.

We now extend Kitaev's model by allowing for the inevitable survival of residual errors in any realistic implementation of a quantum computing machine. In other words, we will extend the analysis to include fault-tolerant operation.

We consider a classical computation for fixed input size, and an implementation
of a quantum computer (appropriately specified using the QCC) that is intended to provide input data for the classical computation. Inspection of the following diagrammatic restatement of the QCC,
\begin{equation} \label{nearly-commutative-diagram}
\begin{diagram}
{\mathbf{T}}(H_\mathrm{comp})  & \rBond^{P} & {\mathbf{T}}(H_\mathrm{comp}) \\
\uBond<{\mathcal{M}_{{\{\mathrm{l}\rightarrow\mathrm{c}\}}}} & \boxed{\alpha}  &  \dBond>{\mathcal{M}_{{\{\mathrm{c}\rightarrow\mathrm{l}\}}}} \\
{\mathbf{T}}(H_\mathrm{logical}) & \rBond_{G} & 
{\mathbf{T}}(H_\mathrm{logical}) \end{diagram}~,
\end{equation}
shows that the completely positive map $P$ acting on arbitrary density matrices, can implement the idealized, perfect quantum computation, $G = U\rho U^{\dag}$, with an inaccuracy no greater than $\alpha$, and hence enables fault-tolerant quantum computation. Recall also that Diagram (\ref{commutative-diagram_mixed}) connects the quantum computation, $G$, to an intended subsequent classical computation, $F$. 
Given this, we show by combining Diagrams (\ref{commutative-diagram_mixed}) and (\ref{nearly-commutative-diagram}), that {\em realistic quantum computation} characterized by residual errors will provide a {\em correct classical computation}, as long as the sum $p+\alpha$ of the probabilistic and implementation inaccuracies is sufficiently small. This is formalized in the following theorem:
\begin{thm} \label{basic-probabilistic-estimate} Suppose Diagrams (\ref{commutative-diagram_mixed}) and (\ref{nearly-commutative-diagram}) hold. Using the compositionality property of these diagrams, proved below, we adjoin Diagram (\ref{nearly-commutative-diagram}) to Diagram (\ref{commutative-diagram_mixed}) to obtain the following diagram: 
\begin{equation} \label{nearly-commutative-diagram-1}
\begin{diagram}
{\mathbf{T}}(H_\mathrm{comp})  & \rBond^{P} & {\mathbf{T}}(H_\mathrm{comp}) \\
\uBond<{\opr{{\tilde I}_\mathrm{m}}} & \boxed{\alpha + p}  &  \dBond>{\opr{{\tilde O}_\mathrm{m}}} \\
X & \rBond_{F} & 
Y \end{diagram}~,
\end{equation}
which is nearly commutative in the sense that
\begin{equation}\label{approximation-cond-for-devices}
\opr{tr}\left(\sqrt{\opr{E}_{F(x)}}P^{\mathcal M}({\opr{{I}_\mathrm{m}}}(x))\sqrt{\opr{E}_{F(x)}}  ~\right) > 1 - (p+\alpha)~,
\end{equation}
where ${\opr{{\tilde I}_\mathrm{m}}}\equiv
{\mathcal{M}_{{\{\mathrm{l}\rightarrow\mathrm{c}\}}}}\circ{\opr{I_\mathrm{m}}}$,
${\opr{{\tilde O}_\mathrm{m}}}\equiv
{\opr{O_\mathrm{m}}}\circ {\mathcal{M}_{{\{\mathrm{c}\rightarrow\mathrm{l}\}}}}$,
and $P^\mathcal{M} \equiv \mathcal{M}_{\{\mathrm{c}\rightarrow\mathrm{l}\}} P \mathcal{M}_{\{\mathrm{l}\rightarrow\mathrm{c}\}}$.
\end{thm}
\begin{proof}
\begin{align*}
\opr{tr}& \left(\sqrt{\opr{E}_{F(x)}}  P^{\mathcal M}(\opr{I}_\mathbf{m}(x)) \sqrt{\opr{E}_{F(x)}}~\right)  \\ 
& = \opr{tr}\left(\sqrt{\opr{E}_{F(x)}}\bigl\{
P^{\mathcal M}(\opr{I}_\mathbf{m}(x)) - U \opr{I}_\mathbf{m}(x) U^\dagger \bigr\} \sqrt{\opr{E}_{F(x)}}~\right)\\
 & + \opr{tr}\left(\sqrt{\opr{E}_{F(x)}} U \opr{I}_\mathbf{m}(x) U^\dagger \sqrt{\opr{E}_{F(x)}}~\right)\\
& >  -\alpha + 1 - p.
\end{align*}
\end{proof}
Diagram (\ref{nearly-commutative-diagram-1}) relates the actual implementation $P$ of the quantum computation to the intended classical computation $F$. 
The above result implies that a {\em quantum} computer realization, $P$, satisfying the QCC and hence operating fault-tolerantly in the presence of residual errors, correctly implements an instance of a {\em classical} probabilistic computation. In the case $Y$ is binary, the classical probabilistic computation succeeds by majority voting if $\alpha + p <1/2$.
Note that in the idealized limit in which error correction {\em perfectly} and {\em permanently} removes all residual errors ({\em i.e.}, in the limit $\alpha = 0$), our result reduces to the corresponding result of the Kitaev model ({\em {i.e.}}, eq.\eqref{approximation-cond-for-devices} reduces to eq.\eqref{approximation-cond}).
This concludes our extension of the Kitaev model.

Elaborating on this result, we see from the
stability result of Proposition~\ref{stability}, that implementation of the
classical computation is stable under small perturbations in the SO norm
of $P$~\cite{f15}. The use of the SO norm in the statement of the QCC allows us to make 
the strongest possible statment of Theorem~\ref{basic-probabilistic-estimate} above.  
By Theorem 9.1 of~\cite{NielsenChuang}, for any trace preserving
positive superoperators $P$ and $P^\prime$,
\begin{equation}
\begin{aligned}
\|P - &{P^\prime}\|_\mathrm{SO}^\mathrm{sa} \\
&= \sup_{E, \rho} \sum_{y \in Y}
{\Big |} \opr{tr} \sqrt{\opr{E}_y} P(\rho) \sqrt{\opr{E}_y} - \opr{tr} \sqrt{\opr{E}_y} {P^\prime}(\rho) \sqrt{\opr{E}_y}{\Big |}
\end{aligned}
\end{equation}
as $E$ varies over POVMs and $\rho$ varies over density matrices.
Since the initialization operators $\opr{{\tilde I}}_\mathbf{m}$
in diagram~\eqref{nearly-commutative-diagram-1} map to arbitrary
density matrices, the SO norm is the smallest norm which can be used
in Theorem~\ref{basic-probabilistic-estimate}.
Thus, the use of the SO norm in the definition of the QCC, 
as opposed to some alternative operator norm, 
plays 
a crucial role in establishing Theorem~\ref{basic-probabilistic-estimate}, 
and hence in quantifying  
how well the quantum computer performs the desired computation.  
In particular, the use of the SO norm 
allows the strongest possible statement of the conditions under which a quantum 
computation is successful.

\section{Distinguishability, Composability and the QCC}

The QCC precisely specifies the constraints that must be satisfied in order for a
candidate quantum computer design to implement a given unitary transformation, up to
a prescribed inaccuracy $\alpha$. The explicit mathematical statement of the QCC makes use of the SO norm to characterize distances between superoperators. Discussions of the distinguishability and composability of 
superoperators in the context of quantum information-theoretic applications~\cite{kitaev1997,aharonov98quantum,watrous}, however, sometimes characterize distances between superoperators using the diamond norm.  The diamond norm is defined as 
follows: 
\begin{dfn}
The diamond norm $\| P \|_\diamond$ of a superoperator $P$ is the supremum of the SO norm of $P \otimes I_n$ for every n. Obviously the diamond norm is at least as large as the SO norm.
\end{dfn}
In this Section we show that physical considerations dictate the use of
the SO norm, and not the diamond norm, in the proper statement of the QCC.
For purposes of analysis, we will designate a variant ``QCC," obtained by replacing 
the SO norm with the diamond norm, 
by the symbol $\mathrm{QCC}_\diamond$, corresponding to the inequality 
\begin{equation}\label{diamond-QCC}
\|  P^\mathcal{M}_Q - G_Q \|_\diamond \leq \alpha~,
\end{equation}
where $G_Q$ is the desired (ideal) quantum computation, 
$P_Q$ is the intended physical implementation, 
and $P^\mathcal{M}_Q \equiv \mathcal{M}_{\{\mathrm{c}\rightarrow\mathrm{l}\}} P_Q \mathcal{M}_{\{\mathrm{l}\rightarrow\mathrm{c}\}}$.  

We consider a quantum computer 
that is initially uncorrelated with 
its environment.  The state of the system is then $\rho_{QE}=\rho_Q \otimes \rho_E$, 
where $Q$ and $E$ denote 
the quantum computer and the environment, 
respectively.
In this case it 
makes no difference whether we state 
the QCC as 
\begin{equation}
\forall \rho_Q~,~\Vert \left(P^\mathcal{M}_Q - G_Q \right) \rho_Q\Vert_1 \leq \alpha~,
\end{equation}
which is equivalent to the proper QCC under the SO norm, or as 
\begin{equation}
\forall \rho_Q \forall \rho_E~,~ \Vert \lbrack \left(P^\mathcal{M}_Q - G_Q \right) \otimes I_E\rbrack 
\rho_Q \otimes \rho_E \Vert_1 \leq \alpha~,
\end{equation}
which is equivalent to the $\mathrm{QCC}_\diamond$ restricted to uncorrelated
states $\rho_Q 
\otimes \rho_E$.  {\em With this restriction understood}, 
the proper QCC, and the variant $\mathrm{QCC}_\diamond$,
furnish equivalent physical statements.

However, it is {\em not} true that the proper QCC and $\mathrm{QCC}_\diamond$
furnish equivalent physical statements in general, since the diamond norm is not restricted to uncorrelated states. If we were to 
use the $\mathrm{QCC}_\diamond$ instead of the proper QCC, we would inevitably reject 
quantum computer implementations that do, in fact, 
give correct results.  To see why this is so, consider a 
completely positive map $P^\mathcal{M}_Q$ that satisfies the proper QCC:
\begin{equation}
\Vert P^\mathcal{M}_Q - G_Q \Vert_\mathrm{SO}^\mathrm{sa} \leq \alpha~.
\end{equation}
It follows from Theorem \ref{basic-probabilistic-estimate} above that the 
implementation $P_Q$ of the quantum computation 
gives the desired results.  $P_Q$ therefore 
furnishes a viable implementation of a quantum computer.  In spite of this, $P_Q$ does 
{\em not} necessarily satisfy the $\mathrm{QCC}_\diamond$.  In general there are 
correlated states $\rho_{QE} \neq \rho_Q \otimes \rho_E$ such that 
\begin{equation}
\Vert \lbrack \left(P^\mathcal{M}_Q - G_Q \right) \otimes I_E\rbrack 
\rho_{QE} \Vert_1  > \alpha ~,
\end{equation}
as pointed out in \cite{aharonov98quantum, watrous}, and so 
\begin{equation}
\Vert  P^\mathcal{M}_Q - G_Q \Vert_\diamond  > \alpha~,
\end{equation}
which violates the $\mathrm{QCC}_\diamond$ ({\em cf}  eq.(\ref{diamond-QCC})).  
Thus, the replacement of the SO norm with the diamond norm in the QCC would result in the 
erroneous rejection  
of an acceptable implementation.  

In other words, the $\mathrm{QCC}_\diamond$ is a {\em sufficient} condition for the successful implementation of a quantum computation, but it is not a {\em necessary} condition. With the SO norm the proper QCC provides both a sufficient and necessary  condition for successful quantum computation~\cite{f16}. The SO norm is clearly the correct norm to use in stating the QCC.  

The diamond norm may nevertheless be useful for certain mathematical applications.
For example, when considering the decomposition of a system into constituent components as a purely mathematical problem, one must consider the fact that, for  
arbitrary superoperators $Q_A$ and $Q_A^\prime$, 
$\Vert \lbrack \left(Q_A - Q_A^\prime \right) \otimes I_B\rbrack \rho_{AB}\Vert_1$  may be 
very large even when $\Vert \left(Q_A - Q_A^\prime \right) \mathrm{Tr}_B \rho_{AB}\Vert_1$ 
is very small \cite{aharonov98quantum, watrous}. 
Mathematical distinguishability in this context  
would call for the use of the diamond norm \cite{kitaev1997}. 

Based on the preceding paragraph, it might appear that physical problems involving the analysis of constituent parts of a quantum computer should be expressed in terms of the $\mathrm{QCC}_\diamond$. However, it is crucially important to note that, due to the definition of the diamond norm, the $\mathrm{QCC}_\diamond$ can provide a rigorously correct description of the dynamics only in the idealized scenario in which the states of the constituents remain {\em fully} uncorrelated from the environment for the duration of the quantum computation. Since correlations with the environment will typically occur in any practical implementation, the $\mathrm{QCC}_\diamond$ cannot characterize the physics in such circumstances. In summary, although the diamond norm provides the correct {\em mathematical} description for the problem of splitting a component into constituent parts, the $\mathrm{QCC}_\diamond$ does not provide the correct {\em physical} solution to the corresponding problem~\cite{f17}.

\section{Conclusion}

We have shown on the basis of the stability properties of the QCC that realistic quantum computation is possible despite variations in system parameters. We then used the QCC to extend Kitaev's model to show that reliable classical computation can be carried out using as input the results of fault-tolerant, but imperfect, quantum computation. We showed that the use of the SO norm in the statement of the QCC plays a critical role in establishing this relationship between quantum and classical computation by providing the strongest possible statement of the accuracy result. We also demonstrated that replacement of the SO norm by the diamond norm leads to the erroneous rejection of acceptable implementations. These properties of the SO norm indicate unambiguously that it is the correct choice for measuring distances between superoperators in the context of the QCC. Finally, we contrasted the QCC problem with the problem of composability of quantum operations. Although the diamond norm provides the correct mathematical description for the problem of splitting a component into constituent parts, we note that the $\mathrm{QCC}_\diamond$ (based on the diamond norm), does not provide the correct physical solution to the corresponding problem since it is incompatible with correlations with the environment.

\section{Acknowledgements} This research was supported under MITRE Technology Program Grant 07MSR205. We would like to thank D Gottesman and D Lidar for helpful comments.

\bibliography{refs}

\begin{thebibliography}{1}
\bibitem{f13}
Note that in \cite{aharonov98quantum}, Kitaev and co-authors discuss gates that may
be subject to noise. However, no specific connection between classical and quantum computation is developed in \cite{aharonov98quantum}. Our treatment provides quantitative, specific relations between the ideally defined quantum computation, $U$, the implementation inaccuracy thereof, $\alpha$, the instability of the actual open quantum system evolution, $P$, and the success probability, $p+\alpha$, of the classical computation, $F$.
\bibitem{f14}
It is important to note that the probabilistic inaccuracy of the classical computation, $p$, and the implementation inaccuracy of the quantum computation, $\alpha$, are independent quantities.
\bibitem{f15}
Since the perturbation $P\rightarrow P^\prime$ results in the replacement $\alpha\rightarrow\alpha^\prime$, stable computation requires that $\alpha^\prime + p < 1/2$.
\bibitem{f16}
The QCC is a necessary condition in the sense that the SO norm is the smallest possible norm for which Theorem \ref{basic-probabilistic-estimate} holds.
\bibitem{f17}
We note that the authors of \cite{aharonov98quantum} make use of the diamond norm of a difference of superoperators to bound the accumulation of errors in a quantum circuit. 
However, since their analysis makes use of the diamond norm, their formulation does not properly account for the general situation in which the constituent states become correlated with the environment.
\end{thebibliography}
\bibliographystyle{hplain}

\end{document}